\documentclass[12pt,preprint]{aastex}

\slugcomment{}

\shorttitle{Broad Line Region in NGC 4051}
\shortauthors{Devereux et al.}

\begin{document}

\title{The Broad Line Region in NGC 4051: An Inflow Illuminated by a 10$^5$ K Accretion Disk}

\author{Nick Devereux  and Emily Heaton}
\affil{Department of Physics, Embry-Riddle Aeronautical University,
         Prescott, AZ 86301}
\email{devereux@erau.edu,heatone@erau.edu}   

\begin{abstract}

Adopting a spherically symmetric steady-state ballistic inflow as the kinematic model for the gas distribution responsible for producing the H${\alpha}$ emission line, and a central black hole (BH) mass of 1.7 ${\times}$ 10$^{6}$ M${_{\sun}}$ determined from prior reverberation mapping, leads to the 
following dimensions for the size of the broad line region (BLR) in NGC 4051; an inner radius ${\sim}$ 3 lt-days and a lower limit to the outer radius ${\sim}$ 475 lt-days. Thus, the previously determined reverberation size for the BLR marks just the inner radius of a much {\it larger} volume of ionized gas. 
The number of ionizing photons required to sustain the H${\alpha}$ emission line luminosity exceeds the number {\it observed} to be available from the 
central AGN by a factor of 3 -- 4. Such a large ionizing deficit can be reconciled if the BLR is ionized by a 10$^5$ K accretion disk that is hidden from direct view by the high opacity of intervening H gas. A new definition is introduced for the {\it ionization parameter} that acknowledges the fact that H opacity significantly attenuates the flux of ionizing photons in the {\it large}, partially ionized, nebula surrounding the AGN. Collectively, the results have important implications for BH masses estimated using reverberation radii and the structure of the BLR inferred from velocity-delay maps.

 \end{abstract}


\section{Introduction}

NGC 4051 is a nearby, 17.9 Mpc\footnotemark \footnotetext{ R.B.Tully 2007, private communication.}, spiral (SBbc) galaxy hosting a narrow line Seyfert 1 nucleus that is time variable in the radio \citep{Ho01}, near-infrared \citep{Sug06}, visible \citep{Den10,Pet00}, UV and X-rays \citep[][and references therein]{Als13,Vas09, Kas04, Kom97}. Consequently, NGC 4051 has been the target of several reverberation mapping campaigns culminating in an estimate for the size of the region emitting the bright Balmer lines of Hydrogen (H) \citep[][and references therein]{Den10,Kas05,She03}. Since this low luminosity active galactic nucleus (LLAGN) is radiating at ${\sim}$ 1\% the Eddington limit \citep{Win12,Vas09} the gas kinematics are dominated by gravity which allows the size of the emitting region to be estimated, independently, by modeling the shape of the Balmer lines, given a kinematic model for the H gas. The purpose of this ${\it Paper}$ is to report two new results. First, if the gas kinematics are dominated by inflow, then the reverberation radius measured for NGC 4051 coincides with the radius that also defines the full width at zero intensity for the H${\alpha}$ emission line. NGC 4051 is the second such LLAGN, after NGC 3227 \citep{Dev13}, to yield this result with potentially important consequences for black hole (BH) mass determinations in these and other time-variable-sub-Eddington AGNs. Second, the bright
H${\alpha}$ line requires a factor of 3 -- 4 more ionizing photons than can be supplied by a simple power law connecting the 
Lyman edge to the X-rays. The ionizing deficit can be alleviated by invoking a ${\sim}$ 10$^5$ K blackbody,
evidence for which was first presented by \cite{Kom97}. Following \cite{Vas09}, we identify the blackbody with emission from an accretion disk. These results are described in more detail in Section 3 and discussed in Section 4 where we
highlight the imperative of including photoionization explicitly in any modeling that attempts to decipher the morphological and
kinematic information conveyed in emission lines. 
Our conclusions concerning the photoionization of the HII region surrounding the AGN in NGC 4051 follow in Section 5. We begin, however, with Section 2 where we acquaint the reader with the
H${\alpha}$ emission line in NGC 4051, as observed with the Space Telescope Imaging Spectrograph (STIS) aboard the {\it Hubble Space Telescope (HST)}.

\section{NGC 4051 H${\alpha}$ Emission Line}

NGC 4051 has been rather neglected as there is only one STIS observation of the H${\alpha}$ emission line in the {\it HST} data archive, obtained  March 12, 2000 with the G750M grating under PID 8228 (Table 1). Two dithered exposures showing the H${\alpha}$ emission line were shifted using the ${\bf STSDAS}$ task ${\bf sshift}$ prior to combining with the ${\bf STSDAS}$ task ${\bf ocrreject}$. Subsequently, the ${\bf STSDAS}$ task ${\bf x1d}$ was used to perform a 7 pixel wide extraction along the slit direction and centered on the nucleus, sampling ${\ge}$ 80\% of the encircled energy for an unresolved point source \citep{Pro10}. NGC 4051 is classified as a narrow line Seyfert 1  \citep[NLS1,][]{Ho97} which, technically, is an oxymoron because Seyfert 1 galaxies have broad lines by definition. The HST spectrum, presented in Figure 1, reveals a bright, single-peaked, but narrow line with a FWHM ${\sim}$ 600 km/s. However, the line exhibits a very broad base with full-width at zero intensity (FWZI)  of at least 4000 km/s. Figure 1 illustrates how the ${\bf STSDAS}$-contributed task, ${\bf specfit}$, was used to model and subtract the two [N II] vacuum wavelength, ${\lambda}$6549.85 {\AA}  and $\lambda$6585.28 {\AA}, emission lines, leaving an unadulterated H${\alpha}$ emission line. Emission line fluxes are reported in Table 2 for the [N II], H${\alpha}$
and vacuum wavelength $\lambda$ 6718.29 {\AA} and $\lambda$ 6732.67 {\AA} [S II] emission lines.

\section{Results}

\subsection{Broad Line Region Ionization} 

The number of ionizing photons produced by the central AGN in NGC 4051 can be estimated using the prescription in \cite{Dev11} where the 912 ${\rm \AA}$ flux is extrapolated from the 1450 ${\rm \AA}$ flux using a UV spectral index ${\alpha}$ = 2 (where {\it f}${_\nu \propto \nu^{-\alpha}}$) reported\footnotemark \footnotetext{Based on the STIS observation of \cite{Col01}.} in \cite{Kas04}. An analysis, by \cite{Vas09}, of contemporaneous UV and X-ray measurements obtained in the year 2001 reveal that the X-rays and consequently the UV-X-ray spectral index, ${\alpha_{\rm ox}}$, varies widely on a monthly timescale with values ranging between 1.17 and 1.51. Using this information one can show that in the {\it high state}, corresponding to ${\alpha_{\rm ox}}$ = 1.17, the central AGN generates 1.5 ${\times}$ 10$^{52}$ ionizing photons/sec whereas in the {\it low state}, corresponding to ${\alpha_{\rm ox}}$ = 1.51, this number decreases to 1.0 ${\times}$ 10$^{52}$ ionizing photons/sec. In contrast, the number of ionizing photons required to generate the H${\alpha}$ emission line luminosity observed in the year 2000 (see Table 3) corresponds to 4.1 ${\times}$ 10$^{52}$ ionizing photons/sec, assuming Case B recombination at a temperature of 10${^4}$ K. Thus, there appears to be rather a large, factor of 3 to 4, discrepancy, otherwise known as an ionizing deficit, between the number of ionizing photons produced by the central AGN and the number required by the ionized gas, as previously suspected by \cite{Kom97}. Ionizing deficits have been noted for other LLAGNs by \citet{Mao98} but, to our knowledge, NGC 4051 is the first NLS1.

Even though NGC 4051 is time variable in the X-rays and extreme UV \citep{Utt00}, it is less so near the 13.6 eV ionization edge where most of the ionizing photons are produced  \citep{Kas04, Als13}. Thus, one can not easily appeal to UV variability to reconcile the ionizing deficit for these non-contemporaneous UV-H${\alpha}$ observations. Some of the broad H${\alpha}$ line emission may be enhanced by collisional excitation; difficult to prove in the absence of contemporaneous H${\beta}$ and H${\gamma}$ line measurements, but in NGC 3227, for example, collisional 
excitation appears to brighten the H${\alpha}$ line emission by only ${\sim}$ 44\% \citep{Dev13}. There is also little evidence for significant Galactic or internal 
extinction based on X-ray observations \citep{Win12,Kom97}. On the other hand, the ionizing deficit measured for NGC 4051 {\it can} be alleviated by invoking a 10$^5$ K blackbody, peaking at 20 eV, which would be obscured from direct observation by the high photoionization cross-section of neutral H. However, tantalizing evidence for the tail of such a blackbody was presented previously by \citet[][see their Figure 5]{Kom97} based on {\it ROSAT}
data and more recently rendered as the emission from an accretion disk by \citet[their Figure 2]{Vas09}. The photoionization requirements of such an accretion disk in relieving the ionizing deficit measured for NGC 4051 is discussed further in Section 4.1.
 
\subsection{Modeling the H${\alpha}$ Emission Line Profile}

Most aficionados believe that AGN activity is fueled by inflowing gas. Concordantly, the distinctly triangular profile shape observed for the H${\alpha}$ emission line in NGC 4051 can be modeled as a steady-state spherically symmetric inflow, illustrated in Figure 2. The inflow model, described previously in \cite{Dev11}, has just two free parameters; ${r_i}$ and ${r_o}$, corresponding to the inner and outer radii of the inflow that is ionized by the central AGN and for which recombination radiation is observed.\footnotemark \footnotetext{Obviously the inflow does not suddenly stop at the inner radius, but the reason for the absence of Balmer recombination radiation inside the inner radius is open to speculation. A likely explanation is that the electron temperature exceeds 10${^4}$ K.} Following the procedure described previously in \cite{Dev13}, values for these two free-parameters that best represent the observed H${\alpha}$ emission line shape, and the associated 1$\sigma$ uncertainties on these parameters, were estimated by calculating reduced Chi-squared (${\chi^2_{red}}$) for a grid of models spanning 9 and 10 values of the inner and outer radii, respectively. ${\chi^2_{red}}$ minimum occurred for ${r_i}$ = 3.5 $^{+2.7}_{-0.1}$ ${\times}$ 10${^4}$ ${r_g}$ 
where $r_g$ = ${GM_{\bullet}/c^2}$; the gravitational radius. Only a lower limit could be obtained for the outer radius because ${\chi^2_{red}}$ asymptotes to a limiting minimum value  at ${r_o}$ ${\ge}$ 4.5 ${\times}$ 10${^6}$ ${r_g}$ as shown in Figure 3. 
Adopting the reverberation BH mass of 1.73 ${\times}$ 10${^6}$ M$_{\odot}$, reported by \cite{Den10}, leads to a physical size for the inner radius, 
${r_i}$ ${\sim}$ (3.4 $^{+2.7}_{-0.1}$) lt-days, which is comparable to the reverberation radius of ($3.0 \pm 1.5$) lt-days reported by \cite{She03},
and ($4.3 \pm 2.1$) lt-days reported by \cite{Kas05}. The inner radius is slightly larger than the reverberation radius ($1.8 \pm 0.5$) lt-days reported most recently by \cite{Den10}. On the other hand, the outer radius of the ionized region corresponds to ${r_o}$ ${\geq}$ 475 lt-days, which is much larger than any measures of the reverberation radius.  Thus, according to our analysis, the reverberation radius marks just the inner radius of a much larger volume of ionized gas. 
NGC 4051 is the second such AGN, after NGC 3227 \citep{Dev13}, to yield this result with important implications for BH mass determinations discussed further in Section 4.2.

\section{Discussion}

Most reverberation mapping studies, to date, have measured the mean time-delay in the correlated response of, usually, the Balmer emission lines to changes in the adjacent continuum, leading to an estimate of the {\it mean radius of the BLR} \citep[e.g.,][and references therein]{Gri13}. However, an often overlooked
fact is that the time-variable component of the emission line, characterized by the parameter F$_{\rm var}$, represents typically only ${\sim}$ 10\% of the total
line flux \citep[e.g.,][]{Pet04}. Thus, reverberation mapping is insensitive to ${\sim}$ 90\% of the line flux that is not variable, at least on the time-scales that these AGNs have been 
monitored. Consequently, one can anticipate that reverberation mapping must be measuring just the {\it inner radius} of a much {\it larger} volume of 
ionized gas and is therefore not the {\it mean radius of the BLR} at all. This anticipated result has now been demonstrated quantitatively for NGC 3227 \citep{Dev13} and here for NGC 4051. In both cases, the reverberation radius consistently coincides with the inner radius of a much larger volume of partially ionized, inflowing gas. A magnified view of the BLR in NGC 4051 is visualized in Figure 4 illustrating the gas and dust reverberation radii measured by \cite{Den10}
and \cite{Sug06}, respectively. Photoionization modeling, described in more detail in the following for NGC 4051, reveals that in both AGNs, the reverberation radius coincides with a transition region; from partially ionized to fully ionized H gas, as the AGN is approached. Collectively, these results have important implications for BH masses estimated using reverberation radii and the structure of the broad line region inferred from velocity-delay maps
as discussed further in Section 4.2.

\subsection{Photoionization Model}

The photoionization model, described previously in \cite{Dev13}, is analogous to the one employed to model an HII region ionized by an O star \citep{Ost06}.
In the case of NGC 4051, following \cite{Vas09}, the source of ionization is
represented by Equation 1 as the superposition of a time-variable power-law component plus a 10$^5$ K blackbody identified with radiation from an accretion disk, alluded to previously in Section 3.1. The relative scaling of the two components is designed to mimic the ${\sim}$ 20\% decrease in the 
H${\alpha}$ emission line luminosity between March, 2000,  when it was observed with {\it HST} and May through July of that same year when it was monitored
by \cite{She03}. The number of ionizing photons per second per unit frequency, $N(\nu)$, is parameterized, in SI units, as

\begin{equation}
 N(\nu)_{ion} = L_{o}  (\nu_o/\nu)^{\alpha} /  (h \nu ) +  (1.1 \times 10^{23} \nu^2 / (c^2 (e^{h \nu /k T} -1)) 
\end{equation}

where T = 10$^5$ K and ${h}$ is Planck's constant. 

In the {\it high}-state, integration of Equation 1 from the photoionization threshold of H, ${\nu_o}$, to 3${\times}$ 10$^{19}$ Hz, yields
4.1 ${\times}$ 10$^{52}$ ionizing photons/sec for $L_{o}$ = 1.77 ${\times}$ 10$^{19}$ W/Hz and ${\alpha}$ = 1.17. The corresponding values for the {\it low}-state are 3.4 ${\times}$ 10$^{52}$ ionizing photons/sec, where $L_{o}$ = 1.51 ${\times}$ 10$^{19}$ W/Hz and ${\alpha}$ = 1.51. 
According to this photoionization model, in the {\it high}-state the 10$^5$ K accretion disk provides ${\sim}$ 80\% of the ionizing photons with the remaining 20\% contributed by the time-variable power-law component.\footnotemark \footnotetext{Interestingly, the H${\alpha}$ line flux exhibits long-term variability on a monthly timescale at the ${\sim}$ 20\% level \citep{Den10,She03} which is tempting to attribute to the time-variable X-ray power-law component. However, explaining the emission line-X-ray variability
of NGC 4051 is not our main purpose here as that task that has challenged several pundits over the years with very little consensus \cite[see][for a quick study]{She03}.}  An appealing attribute of our model is that it explains the observation of \cite{She03} and \cite{Pet00} that the highly variable {\it observable} X-ray continuum does not contribute significantly to the production of the Balmer emission lines.
This is because, in our model, ${\geq}$ 80\% of the ionization required by the Balmer lines
is provided by the hot, ${\sim}$ 20 eV, UV-emitting, accretion disk, which, as noted
previously in Section 3.1, is {\it hidden from direct view} by the high opacity of H at these energies.
Collectively, in our interpretation, the Balmer line emission is produced by a physically large, spherically symmetric, steady state inflow that is photoionized by a much smaller hot, 10$^5$ K, accretion disk. Such a model guarantees a very high, ${\sim}$ 100\%, covering factor for the ionized gas.
Our interpretation is very different from the model of \cite{Pet00} who associate the Balmer line emission {\it directly} with a low inclination, i.e., nearly face-on, accretion disk. 

One would expect the ionized region
in NGC 4051 to respond to the known X-ray variability and indeed the photoionization model shows that when the AGN is in the {\it high}-state\footnotemark \footnotetext{Defined to be 4.1 ${\times}$ 10$^{52}$ ionizing photons/sec.}, the ionization fraction, the electron density and the perceived size of the emitting region all increase compared to when the AGN is in a {\it low}-state\footnotemark \footnotetext{Defined to be 3.4 ${\times}$ 10$^{52}$ ionizing photons/sec.}. These trends are illustrated in Figure 5. Interestingly, the model reveals that the gas is {\it always} completely ionized at the reverberation radius, regardless of which state the AGN is in. The model also reveals that the reverberation radius coincides with a rapid change in the photoionization opacity, ${\tau(\nu,r)}$. Consequently, we introduce a new definition for the ionization parameter,

\begin{equation}
\Gamma(r) =\frac { \int^{\nu_{max}} _{\nu_{o}} N({\nu})_{ion} e^{-\tau(\nu,r)}d{\nu}}{4  \pi r^2 c~ n_e}
\end{equation}

which acknowledges the fact that opacity decreases the flux of ionizing photons with increasing radius, ${r}$, in the {\it large}, partially ionized, nebula
surrounding the AGN. It is significant that this more realistic estimator yields values for ${\Gamma}$ that agree with those quoted for the {\it warm absorber}
seen in the X-rays \citep{Kom97,Win12}. However, since ${\Gamma}$ is predicted to be virtually independent of both radius and the activity state of the
AGN, the {\it warm absorber} could be located anywhere inside the partially ionized volume.
Collectively, the photoionization model reveals
that the reverberation radius marks
a region of transition as the AGN is approached; from high to low opacity and from partially ionized to completely ionized, with a concurrent 
decrease in the ionization parameter. 

Modeling also reveals that the size of the photoionized region depends quite sensitively on the assumed neutral H gas density. Even though the AGN in NGC 4051 produces a similar number of ionizing photons to the one in NGC 3227 \citep{Dev13}, the AGN in NGC 4051 is able to ionize a region that is a factor of ${\sim}$ 4 -- 5 larger in radius, a feat that is only possible if the neutral gas density is about one order of magnitude lower in NGC 4051 than in NGC 3227. An immediate consequence of the larger ionized volume is a narrow Balmer line profile that earned NGC 4051 the NLS1 designation.
Even though the neutral H gas density in the region that the AGN ionizes may be low compared to NGC 3227, the electron density asymptotes to 
a similarly high value, ${\ge}$ 10$^6$ cm$^{-3}$, at the reverberation radius.
Some physical properties of the inflow for NGC 4051 are presented in Table 3 using the prescription described previously in \cite{Dev11}. An important result is
that the upper limit to the mass inflow rate of ionized gas, corresponding to ${\sim 6 \times10^{-3}~\rm M_\odot~yr^{-1}}$, is not particularly unusual compared to other LLAGNs, but it is about a factor of 3 less than required to explain the observed 2 -- 10 keV X-ray luminosity of NGC 4051 in terms of radiatively inefficient accretion. 

Since there is no evidence, nor any reason, to expect the ionizing capabilities or the gas density to be the same 
for all AGNs, the outer radius of the ionized region is expected to vary widely from one AGN to the next and consequently, the emission line profile shapes to
vary from one AGN to the next. This observation highlights the imperative of including the entire Balmer emission line, not just the time-variable component,
when using broad emission line profiles to infer the {\it shape} and {\it size} of the BLR.

\subsection{Virial Black Hole Masses}

In the context of the inflow model, the inner radius, synonymous now with the reverberation radius, defines the full width at zero intensity of the 
broad H${\alpha}$ emission line as described previously in \cite{Dev13}. Consequently, the correct virial product for computing the BH mass, at least for NGC 4051, is 

\begin{equation}
M_{\bullet} = R {\Delta}V_{HWZI}^{2}/2G
\end{equation}

where {\it R} is the reverberation radius  (3.4 lt-days) and  ${\Delta}V_{HWZI}$ is the half width at zero intensity (HWZI) of the emission line (2250 km/s). Using these numbers in Equation 3 immediately recovers the reverberation BH mass, {\it M$_{\bullet}$} = 1.7 ${\times}$ 10${^6}$ M$_{\odot}$, cited by \cite{Den10}. 
This approach is to be contrasted with the procedure advocated by \cite{Pet04} in which a virial product resulting from the reverberation radius and the line dispersion, ${\sigma_{line}}$, squared, is multiplied by a factor {\it f} = 5.5 to yield a BH mass according to the relation

\begin{equation}
M_{\bullet} =  f R {\sigma_{line}}^{2}/G
\end{equation}

The factor of 5.5 is the empirically determined value that 
scales the virial products to BH masses, {\it M$_{\bullet}$}, so that the latter are on the same {\it M$_{\bullet}$} -- ${\sigma}_*$ relation as those measured directly for other galaxies using gas and star kinematics \citep{Onk04, Pet04}. In theory, the parameter, {\it f}, in Equation 4, depends on the kinematic model adopted for the BLR gas.
However, the procedure described by \cite{Pet04} for computing BH masses has evolved in lieu of a specific kinematic model. The main point we are promoting in this ${\it Paper}$ is that the BLR gas ${\it is}$ consistent with a specific kinematic model, namely ${\it free-fall}$, leading to a more direct route to computing BH masses using reverberation radii in Equation 3.

\section{Conclusions}

High resolution spectroscopy of the LLAGN in NGC 4051 obtained with {\it HST} reveals a distinctly triangular profile shape for the H${\alpha}$ emission line which has been modeled as a steady-state spherically symmetric ballistic inflow. According to this interpretation our principal conclusions are 1). The inner radius of the region producing the Balmer line emission is ${\sim}$ 3 lt-days and consistently
coincides with previous measures of the reverberation radius. 2). The outer radius of the region producing the Balmer line emission is at least 475 lt-days. Thus, the  reverberation radius measured for NGC 4051 locates just the inner radius of a much larger volume of partially ionized,
inflowing, H gas, the full extent of which is unlikely to be revealed by velocity-time mapping. 3). If the Balmer lines are produced by an inflow then the correct virial product for computing the BH mass should utilize the reverberation radius in
conjunction with the half-width at zero-intensity of the emission line.  4). Photoionization modeling reveals that the reverberation radius
coincides with a transition from partially ionized to fully ionized gas as the AGN is approached and coincides with a concurrent decrease in the photoionization opacity, independent of whether the LLAGN is in a {\it high} or {\it low} ionization state.
5). The rather large factor of 3 -- 4 ionizing deficit observed for the LLAGN is mitigated if ${\sim}$ 80\% of the ionization is provided by a 10$^5$ K blackbody, identified with a UV emitting accretion disk, that is hidden from direct view by the high opacity of intervening H gas. 6). A new definition for the {\it ionization parameter}, ${\Gamma}$(r), acknowledges the significant attenuation of the ionizing photon flux in the {\it large} volume of partially ionized gas responsible for the Balmer line emission. This new definition yields values for ${\Gamma}$ that concur with those reported in the published literature for the {\it warm absorber} 
seen in the X-rays.  

\acknowledgments

The authors than the referee for useful comments that improved the presentation of the Paper.

{\it Facilities:}  \facility{HST (STIS)}

\clearpage

\begin{figure}
\epsscale{1.0}
\begin{center}
\plotone{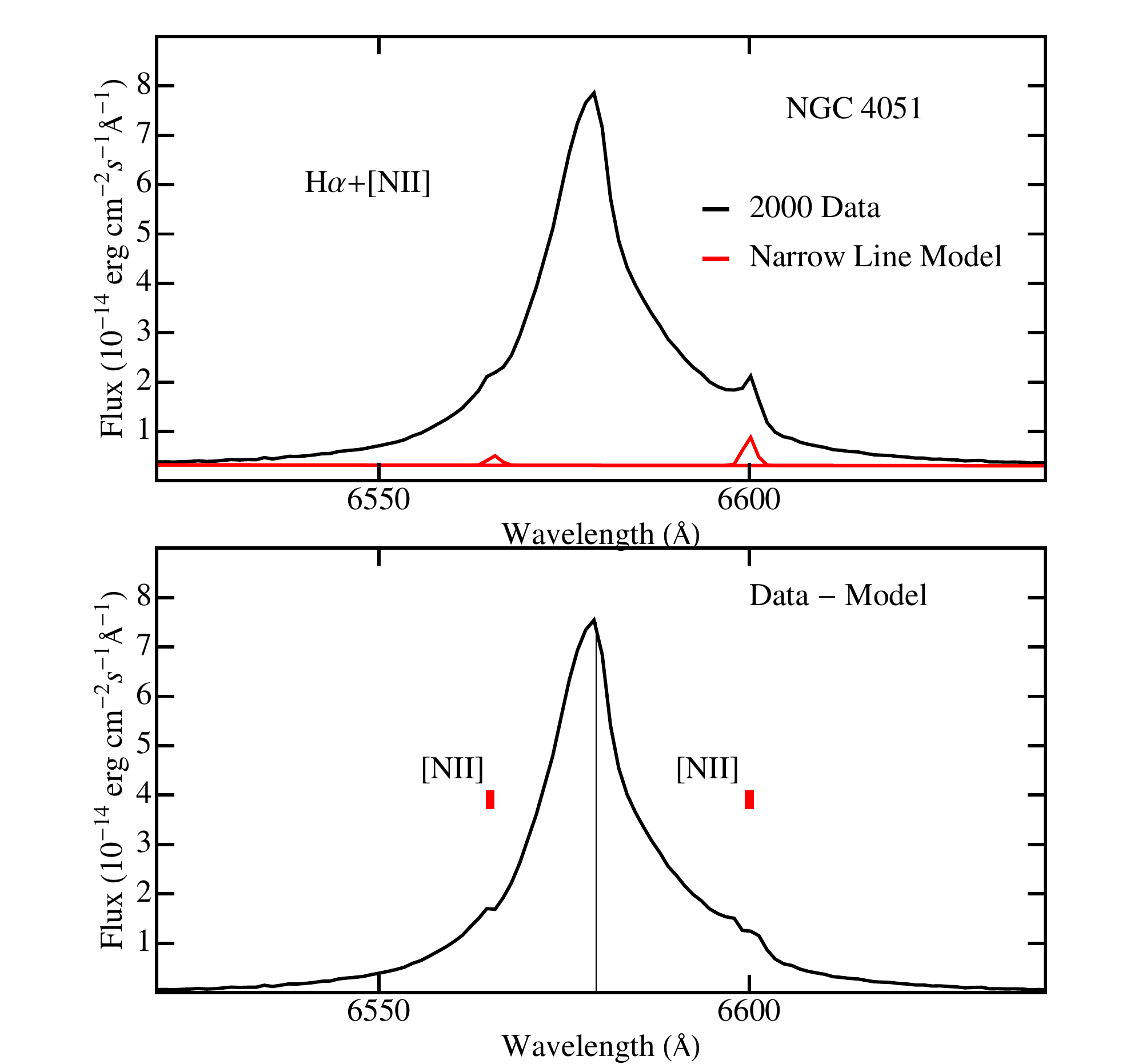}
\caption{H$\alpha$ emission line in NGC 4051.  Top panel: the observed spectrum is shown in black and a model for the forbidden [N II] lines is shown in red (see also Table 2).  Lower panel: The broad H$\alpha$ emission line profile after the forbidden lines have been subtracted.  The central wavelengths of the subtracted lines are indicated in red.  The vertical black line corresponds to the observed (redshifted) central wavelength of H$\alpha$ line.}
\label{default}
\end{center}
\end{figure}

\clearpage

\begin{figure}
\epsscale{1.0}
\begin{center}
\plotone{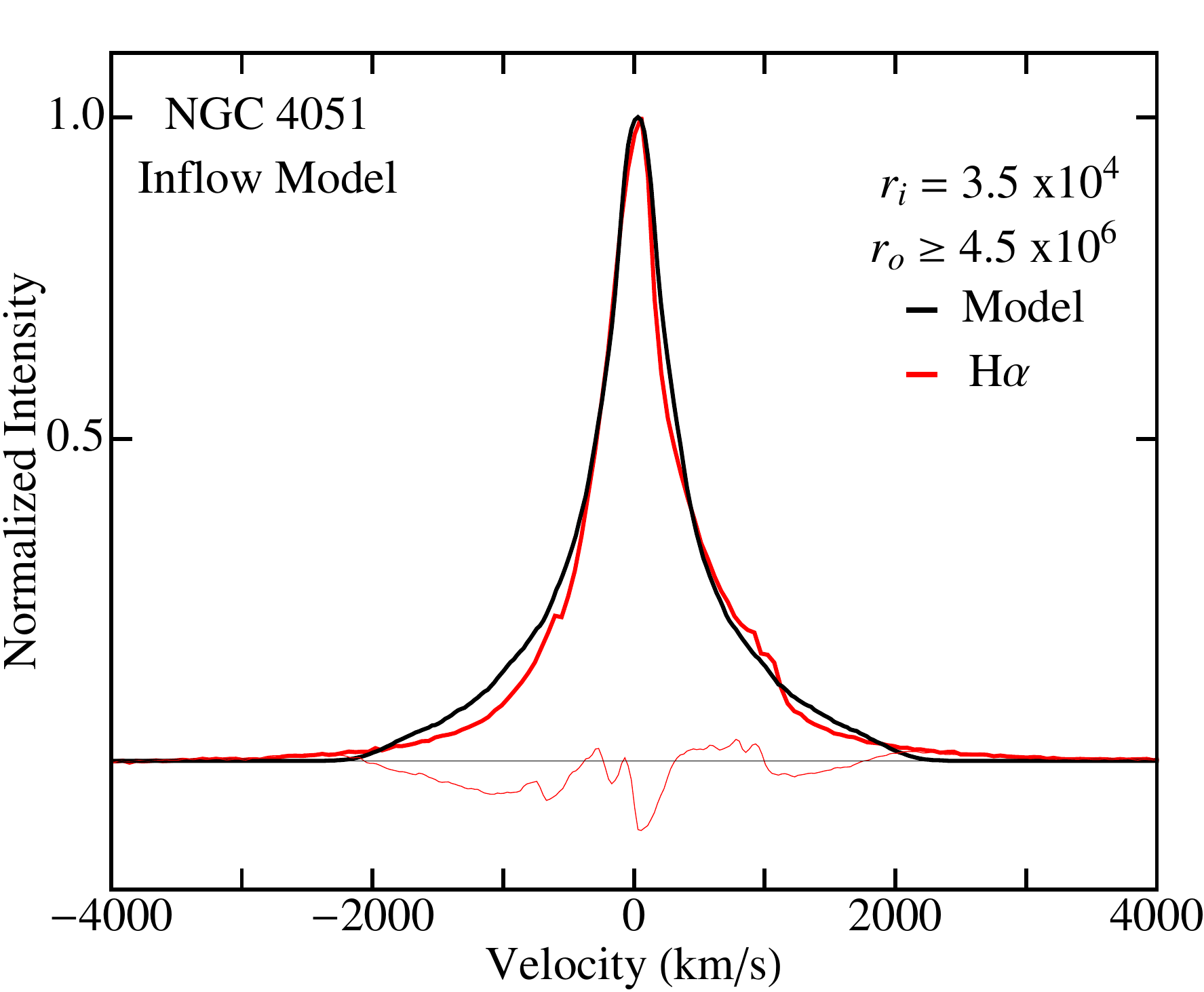}
\caption{Model representation of the H$\alpha$ line in terms of a spherically symmetric inflow. The observed H$\alpha$ emission line is shown in red. The inflow model is shown in black. Residuals are plotted as the thin red line. $r_o$ and $r_i$ are the inner and outer radii of the model inflow expressed in units of the gravitational radius, ${r_g}$. }
\label{default}  
\end{center}
\end{figure}

\clearpage

\begin{figure}
\epsscale{1.0}
\begin{center}
\plotone{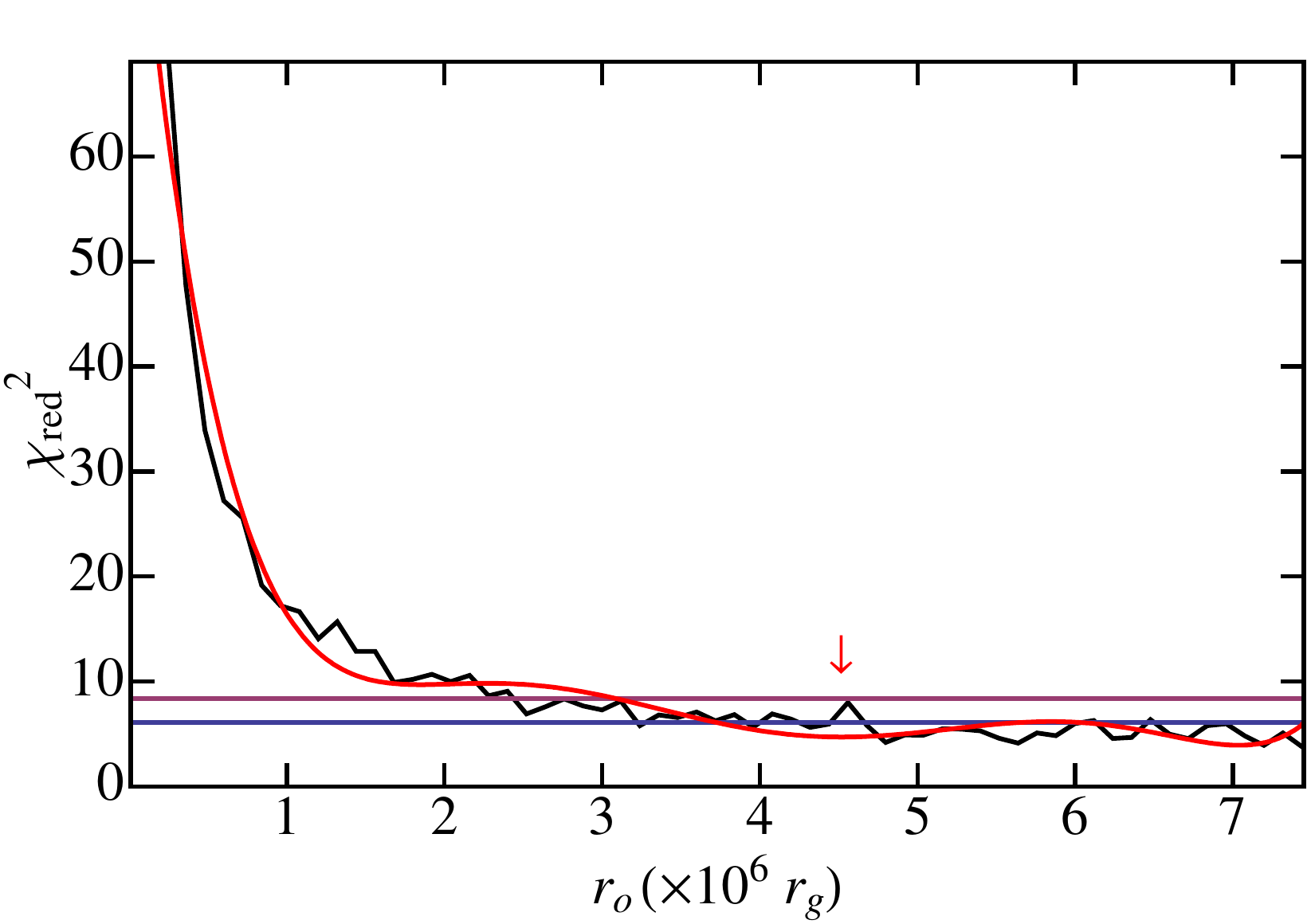}
\caption{The black line represents $\chi_{red} ^2$ calculated for $2.4 \times10^{5} r_g$ ${\le}$ $r_o$ ${\le}$ $7.4 \times10^{6} r_g$, keeping $r_i$ fixed at 3.5 $\times10^{4} r_g$. The two horizontal lines identify the 1$\sigma$ and 2$\sigma$ 
bounds on $\chi_{min} ^2$ = 4. The red line is a 6$^{th}$ order polynomial representation of $\chi_{red} ^2$ for which the first root
that minimizes $\chi_{red} ^2$ to be within 2$\sigma$ of $\chi_{min} ^2$ corresponds to $r_o = 4.5 \times10^{6} r_g$, indicated by the arrow. }
\label{default}
\end{center}
\end{figure}

\clearpage

\begin{figure}
\epsscale{1.0}
\begin{center}
\plotone{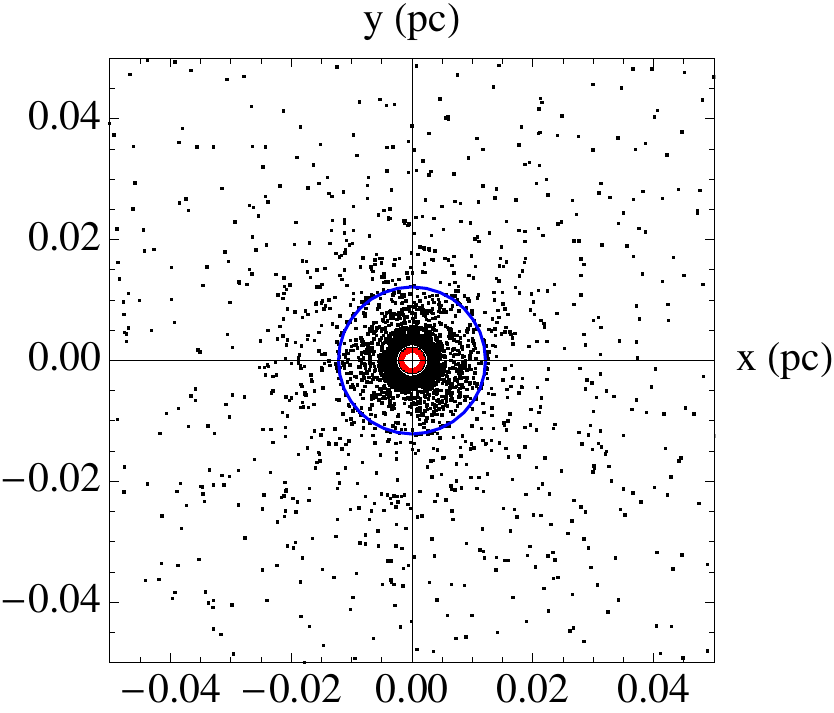}
\caption{Visualization of the model inflow. The red dots represent 7\% of the total number of points employed to define the model emission line, shown in Figure 2, that is observed to be time-variable according to the parameter, ${F\rm_{var}}$, cited in \cite{Den10}. The blue ring represents the mean dust reverberation radius \citep{Sug06}. Dimensions are in units of pc.}
\label{default}
\end{center}
\end{figure}

\clearpage

\begin{figure}
\epsscale{1.0}
\begin{center}
\plotone{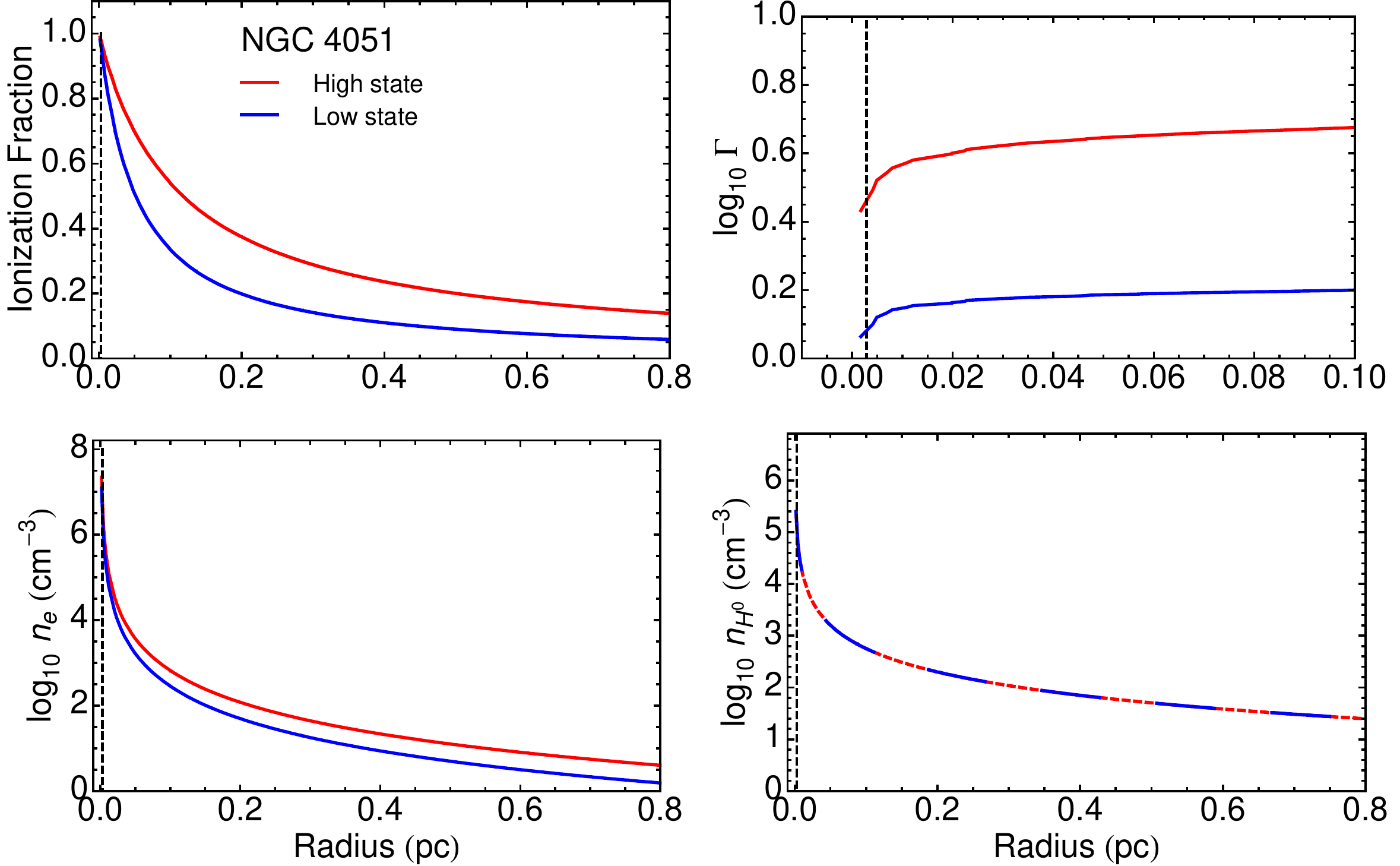}
\caption{Photoionization model results for NGC 4051 illustrating the radial dependence of the ionization fraction, ${\sim r^{-1/3}}$ (Top left panel), the ionization parameter, ${\sim r^{0}}$
(Top right panel), the electron number density, ${\sim r^{-5/2}}$ (Lower left panel) and the neutral H number density, ${\sim r^{-3/2}}$ (Lower right panel). The vertical dashed line identifies the inner radius of the inflow, $r{_i}$, in units of pc (see Section 3.2). }
\label{default}
\end{center}
\end{figure}

\clearpage

\begin{deluxetable}{ccccccccc}
\tabletypesize{\scriptsize}
\tablenum{1}
\tablewidth{0pt}
\tablecaption{NGC 4051 Spectral Datasets}
\tablehead{
\colhead{PID}  & \colhead{Observation Date}  &
\colhead{Grating}    & \colhead{Spectral Range}  &
\colhead{slit}   & \colhead{Dispersion}  &
\colhead{Plate Scale}  & \colhead{Integration Time}  &
\colhead{Data Set} \nl
\colhead{}  & \colhead{}  & \colhead{}    & \colhead{\AA}  &
\colhead{arc sec}   & \colhead{\AA/pixel}  & \colhead{are sec/pixel}  & \colhead{s}  &
\colhead{} \nl
\colhead{(1)}  & \colhead{(2)}  & \colhead{(3)}    & \colhead{(4)}  &
\colhead{(5)}   & \colhead{(6)}  & \colhead{(7)}  & \colhead{(8)}  &
\colhead{(9)}
}
\startdata
8228& 2000 Mar 12& G750M & 6482.0-7054.0& 52 x 0.2& 0.56 & 0.05  & 432 & 05H730030   \nl
8228& 2000 Mar 12& G750M & 6482.0-7054.0& 52 x 0.2& 0.56 & 0.05 & 432 & 05H730040   \nl
\enddata
\end{deluxetable}

\clearpage

\begin{deluxetable}{cccc}
\tabletypesize{\scriptsize}
\tablenum{2}
\tablecaption{Emission Line Parameters for the G750M Nuclear Spectrum Obtained 2000 March 12\tablenotemark{a}}
\tablewidth{0pt}
\tablehead{
\colhead{Line} & \colhead{Central Wavelength\tablenotemark{b}} & \colhead{Flux\tablenotemark{c}} & \colhead{FWHM}   \\
\colhead{} & \colhead{\AA} &  \colhead{10$^{-15}$ erg cm$^{-2}$ s$^{-1}$} & \colhead{kms$^{-1}$} \\
\colhead{(1)} & \colhead{(2)} &  \colhead{(3)} & \colhead{(4)} \\
}
\startdata
$\textrm{[N II]}$ & 6600 $\pm$ 2 & 12\tablenotemark{d} & 88  \\
$\textrm{[N II]}$ & 6563 $\pm$ 2 & 4 & 88  \\
H$\alpha$ (broad) &6579 &1543 $\pm$ 7  &610 $\pm$ 60   \nl
$\textrm{[S II]}$\tablenotemark{e} & 6734 $\pm$ 1 & 11.5 $\pm$ 0.7 & \nodata  \nl
$\textrm{[S II]}$ & 6748 $\pm$ 1 & 9.1 $\pm$ 0.8 & \nodata  \nl
\enddata
\tablenotetext{a}{Table entries that do not include uncertainties are fixed parameters.}
\tablenotetext{b}{Observed wavelength}
\tablenotetext{c}{Measured within
a 0.2{\arcsec}  x 0.35{\arcsec}  aperture. Continuum subtracted but not corrected for dust extinction. Model dependent systematic uncertainties introduce an additional ${\sim}$3\% error not reported in the Table.}
\tablenotetext{d} {The [N II] emission line flux is chosen so as to not over-subtract the broad H${\alpha}$ emission line profile. }
\tablenotetext{e}{Both [S II] lines are skewed which prevents an estimate of their FWHM.}
\end{deluxetable}

\clearpage

\begin{deluxetable}{cccc}
\tabletypesize{\scriptsize}
\tablenum{3}
\tablecaption{Physical Properties of the Inflow}
\tablewidth{0pt}
\tablehead{
\colhead{Parameter} & \colhead{Value} \\
\colhead{(1)} & \colhead{(2)}  \\
}
\startdata
Electron number density, ${\it n{_e}}$, at the inner radius & ${\ge}$  10$^6$ cm$^{-3}$ \\ 
H${{\alpha}}$ luminosity\tablenotemark{a}, $L (H{{\alpha}}$) & 1.46 x 10${^7}$ L${_{\sun}}$ \\
Mass of ionized gas in BLR, $M_{emitting}$ & ${\le}$ 134 M${_{\sun}}$ \\
Volume filling factor of ionized gas in BLR, ${ \epsilon}$ & 2.6 x 10${^{-2}}$ ${\le}$ ${ \epsilon}$ ${\le}$ 1 \\
Ionized gas mass inflow rate\tablenotemark{b}, ${\dot{m}}$ & ${\leq 5.7 \times 10^{-3} ~\rm M_\odot~yr^{-1}}$ \\
2-10 keV X-ray luminosity\tablenotemark{c}, $L_{2-10~keV}$ & $6.3\times 10^{40}~{\rm erg~s^{-1}}$ \\
Mass inflow rate required by AGN, ${\dot{m}}$ & ${\sim 1.6 \times 10^{-2} ~\rm M_\odot~yr^{-1}}$ \\

\enddata
\tablenotetext{a}{measured in the year 2000 (Table 2)}
\tablenotetext{b}{assuming ${ \epsilon}$ $\leq$ 1, at the inner radius.}
\tablenotetext{c}{measured in the year 2001 \citep{Vas09}}

\end{deluxetable}

\end{document}